\documentclass[conference,letterpaper]{IEEEtran}

\IEEEoverridecommandlockouts
\usepackage{bbm}
\usepackage{lineno}
\usepackage{amssymb}
\usepackage{amsmath}
\usepackage{amsthm}
\usepackage{epsfig}
\usepackage{graphicx}
\usepackage{graphics}
\usepackage{float}
\usepackage{subfigure}
\usepackage{multirow}
\usepackage{color}
\usepackage{makeidx}
\usepackage{xspace}
\usepackage{wrapfig}
\usepackage{lipsum}
\usepackage{mathtools}
\usepackage{cuted}
\usepackage{cite}
\usepackage{amsfonts}
\usepackage{textcomp}
\usepackage{enumerate}
\usepackage{widetext}

\usepackage{textcomp}
\usepackage{enumerate}
\usepackage{algpseudocode}
\usepackage{algorithm}

\usepackage[dvipsnames]{xcolor}

\makeindex
\theoremstyle{plain}

\newtheorem{Theorem}{Theorem}
\newtheorem{Proposition}{Proposition}
\theoremstyle{definition}

\newtheorem{remark}{Remark}
\newtheorem{Proof of Lemma}{Proof of Lemma}

\makeatletter

\renewcommand\@endtheorem{\vvv@endmarker\endtrivlist\@endpefalse}
\newcommand\vvv@endmarker{%
  {\unskip\nobreak\hfil\penalty50
  \hskip2em\vadjust{}\nobreak\hfil\openbox
  \parfillskip=0pt \finalhyphendemerits=0 \par
  \penalty 10000 \parskip=0pt\noindent}\ignorespaces}
\makeatother

\definecolor{darkred}{rgb}{1, 0.1, 0.3}
\definecolor{darkblue}{rgb}{0.1, 0.1, 1}
\definecolor{darkgreen}{rgb}{0,0.6,0.5}

\def \T {\mathcal{T}}

\def \W {\mathcal{S}}

\def \F {\mathbb{F}}

\def \supp {\text{supp}}

\allowdisplaybreaks

\def\BibTeX{{\rm B\kern-.05em{\sc i\kern-.025em b}\kern-.08em
    T\kern-.1667em\lower.7ex\hbox{E}\kern-.125emX}}

\cleardoublepage
\begin{document}
    \title{Perfect Multi-User Distributed Computing}
          \author{\IEEEauthorblockN{Ali Khalesi and Petros Elia}
     \thanks{{
     This work was supported by the European Research Council (ERC)
through the EU Horizon 2020 Research and Innovation Program under Grant
725929 (Project DUALITY), ERC-PoC project LIGHT (Grant 101101031), as well as by the Huawei France-funded Chair towards
Future Wireless Networks. The authors are with the Communication Systems Department at
EURECOM, 450 Route des Chappes, 06410 Sophia Antipolis, France (email:
khalesi@eurecom.fr; elia@eurecom.fr).
}}}
\maketitle
\begin{abstract}
     In this paper, we investigate the problem of multi-user linearly decomposable function computation, where $N$ servers help compute functions for $K$ users, and where each such function can be expressed as a linear combination of $L$ basis subfunctions. The process begins with each server computing some of the subfunctions, then broadcasting a linear combination of its computed outputs to a selected group of users, and finally having each user linearly combine its received data to recover its function. As it has become recently known, this problem can be translated into a matrix decomposition problem $\mathbf{F}=\mathbf{D}\mathbf{E}$, where $\mathbf{F} \in \mathbf{GF}(q)^{K \times L}$ describes the coefficients that define the users' demands, where $\mathbf{E} \in \mathbf{GF}(q)^{N \times L}$ describes which subfunction each server computes and how it combines the computed outputs, and where $\mathbf{D} \in \mathbf{GF}(q)^{K \times N}$ describes which servers each user receives data from and how it combines this data. 
     
     Our interest here is in reducing the total number of subfunction computations across the servers (cumulative computational cost), as well as the worst-case load which can be a measure of computational delay. Our contribution consists of novel bounds on the two computing costs, where these bounds are linked here to the covering and packing radius of classical codes. One of our findings is that in certain cases, our distributed computing problem --- and by extension our matrix decomposition problem --- is treated optimally when $\mathbf{F}$ is decomposed into a parity check matrix $\mathbf{D}$ of a perfect code, and a matrix $\mathbf{E}$ which has as columns the coset leaders of this same code. 
\end{abstract}

\section{Introduction}
Distributed computing systems such as MapReduce \cite{dean2008mapreduce} and  Spark \cite{zaharia2010spark} form the backbone for processing computationally hard functions in a variety of real-life applications such as scientific simulations, gaming, as well as training large-scale machine learning algorithms and deep neural networks with high data complexity (cf.~\cite{verbraeken2020survey}). For many such applications, distributed parallel processing techniques become crucial for offloading computations to a group of distributed servers in order to reduce the computation time. 
 
It is the case though that this desired parallelization to the distributed servers, constantly presents new challenges that inspire various works, within the disciplines of information theory and coding theory, seeking to design algorithms and fundamental performance limits.  Such research can be found of computational accuracy~\cite{jahani2021codedsketch,wang2021price,ozfatura2021coded,wan2022cache}, latency and straggler mitigation \cite{raviv2020gradient,reisizadeh2019tree,wan2021distributed,tandon2017gradient}, scalability \cite{li2018near,soleymani2021analog,8437333,8051074}, security and privacy \cite{soleymani2020distributed,khalesi2021capacity,akbari2021secure,nodehi2018limited,wan2022secure}, as well as in the context of communication and computation complexity~\cite{li2018compressed,yu2017optimally,yu2017polynomial2,li2021flexible,jamali2019coded,suh2017matrix,wan2021tradeoff, Brunero1, Parrinello1, Brunero2}.
\nocite{verbraeken2020survey,ulukus2022private}
 
Motivated by the same need to efficiently parallelize multiple computational tasks, our work here studies the known multi-user linearly decomposable scenario introduced in \cite{khalesi2022multi,khalesi2022multi2}, which can be seen as a multi-user extension to \cite{kai1}, and which entails a master node that acts as a total trusted authority in managing $N$ server nodes, serving $K $ users that each demand their function to be computed.  Under the linearly-decomposable assumption, where each function is a linear combination of $L$ basis subfunctions, it was recently shown (cf. \cite{khalesi2022multi,khalesi2022multi2}) that the multi-user distributed computing problem is mathematically equivalent to a sparse matrix factorization problem of the form $\mathbf{F} = \mathbf{D}\mathbf{E}$, where $\mathbf{F} \in \mathbf{GF}(q)^{K \times L}$ describes the linear coefficients of the demands of the users, where the so-called computation and encoding matrix $\mathbf{E} \in \mathbf{GF}(q)^{ N \times L}$ describes which subfunctions each server evaluates and how each server combines the subfunctions' outputs, while the so-called {communication and decoding matrix} $\mathbf{D} \in \mathbf{GF}(q)^{K \times N}$ describes the server-to-user activated connectivity and the manner with which each user combines its received data.

This work here studies the above scenario by jointly considering both the cumulative computational cost across the servers, as well as (under a uniformity assumption in the computational delay of evaluating each subfunction) the worst-case computational load which, as we will discuss later, captures the concept of computational delay. Our main contribution is to connect the above two metrics of our distributed computing problem, to the coding-theoretic concepts of the covering radius and the packing radius respectively. Then, in terms of the construction of schemes that efficiently resolve our distributed computing problem, our work provides a never-before-seen connection between distributed computing and the powerful structure of perfect codes. Deviating from the approach in \cite{khalesi2022multi} which uses covering codes to reduce the computation cost in asymptotic settings, we here show how perfect codes --- which optimize both the covering radius and packing density of codes --- yield an improved solution to our distributed computing problem, both in terms of cumulative as well as worst-case costs, and do so for finite dimensions. To the best of our understanding, this is the first time that perfect codes (and the closely related quasi-perfect codes) have been associated with distributed computing and the equivalent problem of matrix factorization.  
The derived novel bounds on the cumulative computational cost as well as on the computational delay of a multi-user linearly-decomposable system capture the importance of the packing density as well as the packing and covering radius (defined in \cite{etzion2022perfect}) of a code whose parity-check matrix is our communication-and-computing matrix $\mathbf{D}$.
. 
\\

 \noindent{\bf Paper Organization and Notation:}
Section~\ref{System-Model} introduces the system model, Section \ref{Formulating} formulates the problem,  Section~\ref{results} presents the main results, and finally,  Section~\ref{conclusion} concludes. 
In terms of notation, for some $n\in \mathbb{N}$, we define $[n] \triangleq \{1,2,\hdots , n\}$, while for some vector $\mathbf{x}$ or matrix $\mathbf{X}$, we will use $\omega(\mathbf{x})$ (resp. $\omega(\mathbf{X})$) to denote the corresponding Hamming weight. Furthermore, we will use the short-hand notation $\mathbb{F}$ to represent a finite field $\mathbf{GF}{(q)}$. For some vector $\mathbf{x} \in \F^{n}$ and some code $\mathcal{C} \subseteq \F^{n}$, we will use  $d(\mathbf{x},\mathcal{C})$ to represent the Hamming distance of $\mathbf{x}$ to the nearest codeword in $\mathcal{C}$.  For some matrix $\mathbf{H}$, we will use $\mathcal{C}_{\mathbf{H}}$ to represent the linear code whose parity-check matrix is $\mathbf{H}$. Similarly, when we write $\mathbf{H}_{\mathcal{C}}$, we will refer to the parity-check matrix of a linear code $\mathcal{C}$. For some $k\leq n,\: k,n \in \mathbb{N}$, we will also use $\mathcal{C}(k,n)$ to represent a linear code of message length $k$ and codeword length $n$.  Finally, we will use $\supp(\mathbf{x}^{})$ to represent the support of some vector $\mathbf{x}^{} \in \F^{n}$, corresponding to the set of indices of non-zero elements of that vector.

\section{System Model}\label{System-Model}
We consider the multi-user linearly-decomposable distributed computation setting (cf.~Fig.~\ref{Fig: System Model}), which consists of $K$ users/clients, $N$ active (non-idle) servers, and a master node that coordinates servers and users. The process starts with each user asking for a (generally non-linear) function from a space of linearly decomposable functions, where each such function takes several subfunctions as input\footnote{We here implicitly emphasize on scenarios where the computation of these subfunctions far outweighs and subsequent encoding and decoding operations}.
In particular, each desired function can be decomposed into a different linear combination of individual subfunctions $f_{l}(.)$, and thus the demanded function $F_k$ of each user $k \in[K]$, takes the general linearly-decomposable form
\begin{align}
    F_k(.)&\triangleq f_{k,1}f_{1}(.) + f_{k,2}f_{2}(.) \notag+\hdots +f_{k,L}f_{L}(.),\:\: k \in [K]\\
    & =f_{k,1}W_1 + f_{k,2}W_2  +\hdots +f_{k,L}W_L,\:\: k \in [K]\label{DefinitionOfLSFunctions}
\end{align}
where $W_l = f_{l}(.) \in \F ,\: l \in [L]$ is a so-called `file' output, and  $f_{k,l} \in \F ,\: k \in [K], l \in [L]$ are the corresponding linear combination coefficients.

\begin{subsection}{Three Phases of the Distributed Computing Problem}
The model involves three phases, with the first being the \emph{demand phase}, followed by the  \emph{computation and encoding phase}, and then the \emph{communication and decoding phase}. In the {demand phase},  each user $k \in [K]$ sends the information of its desired function $F_k(.)$ to the master node, which then evaluates the linear decomposition of each function as in~\eqref{DefinitionOfLSFunctions}.
Then based on these $K$ desired functions, during the computation and encoding phase, the master assigns some of the subfunctions to each server, who then proceeds to compute these and produce the corresponding files $W_l = f_{l}(.)$. In particular, each subfunction $f_{l}(.), l \in \W_{n} \subseteq [L]$ will be assigned to server $n$ to be computed, for some carefully selected subset $\W_{n} \subseteq [L]$.

During the communication and decoding phase, each server $n \in [N]$ broadcasts in one shot, its own linear combination of the locally computed output files, to a specific\footnote{We here emphasize that the activated topology is not fixed before the demand phase, and that indeed the choice of receiving users for each server a function of the demanded jobs.} subset of users $\T_{n}$. In particular, each server $n$ transmits 
  \begin{align}
  z_{n}\triangleq \sum_{l \in [L]} e_{n,l} W_l,\:\: n\in [N] \label{EncodedFiles}
  \end{align}
   where the so-called encoding coefficients $e_{n,l}\in \F$ are determined by the master.
 Finally during the decoding part, each user $k$ linearly combines the received signals as follows
\begin{align}
    F'_{k} \triangleq \sum_{n \in [N]} d_{k,n} z_{n}\label{DecedFiles}
\end{align}
 for some decoding coefficients $d_{k,n} \in \F, n\in [N]$, determined by the master node. Naturally $d_{k,n} =0,\forall k \notin \mathcal{T}_{n}$.  The functions are successfully evaluated when $F'_k = F_k, \forall k\in[K]$.   
  \begin{figure}
      \centering
      \includegraphics[scale=0.5]{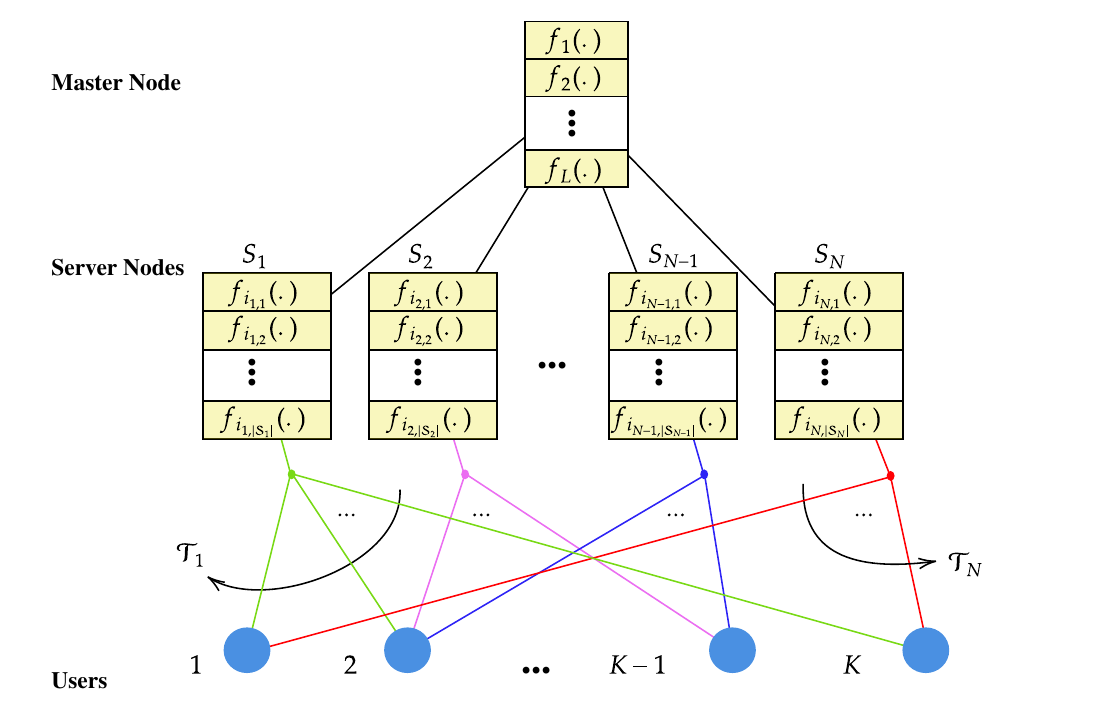}
      \caption{The $K$-user, $N$-server, linearly-decomposable computation setting. After each user informs the master of its desired function $F_k(.)$, each server $n \in [N]$  computes  a  subfunction $W_l = f_{l}(.)$  in $\W_n \subset [L]$. Afterwards, server $n$ broadcasts a linear combination ${z}_{n}$ (of the locally available computed files) to all users in $\T_{n} \subseteq [K]$. This combination is defined by the coefficients $e_{n,l}$. Finally, based on some decoding-coefficient vectors (based on some decoding-coefficient vectors, to be described later on), each user $k \in [K]$ linearly combines (based on decoding vectors $\mathbf{d}_k$) all the received signals. Decoding must produce for each user its desired function $F_k(.)$.}
      \label{Fig: System Model}
  \end{figure}

  \end{subsection}

 \begin{subsection}{Cumulative Computation Cost and Computational Delay}

Recalling that $|\W_{n}|$ indicates the number of subfunctions that server $n$ computes, we consider the \emph{Cumulative Computation Cost}  to take the form
  \begin{align}
      \Gamma \triangleq \sum^{N}_{n=1}|\W_{n}|\label{ComputationCost}
  \end{align}
representing the cumulative number of sub-function computations across all servers. 
   \end{subsection}

Furthermore, assuming that the jobs at each server are computed sequentially --- and under a uniformity assumption that evaluating each file $W_l, l\in [L]$ from the subfunction $f_l(.)$
requires a normalized unit of time --- we may consider a server's computational delay $T_n = |\W_n| $
and thus, under some basic synchronization assumptions, we may consider an overall computational delay that takes the form 
   \begin{align}
       \Lambda \triangleq \max |\W_n|.
   \end{align}
We wish to provide schemes that correctly compute the desired functions, with reduced costs $\Gamma$ and $\Lambda$.

\section{Problem Formulation: One-Shot Setting}\label{Formulating}
To formulate the problem, we use 
\begin{align}
      \mathbf{f}&\triangleq [F_1,F_2,\hdots,F_K]^{\intercal}, \\
      \mathbf{f}_k &\triangleq [f_{k,1},f_{k,2},\hdots,f_{k,L}]^{\intercal},\: k \in [K],\label{function-vectors-1}\\
    \mathbf{w}&\triangleq [W_{1},W_{2},\hdots,W_{L}]^{\intercal}\label{message-vectors-1}
\end{align}
where $\mathbf{f}$ represents the vector of the output demanded functions (cf.~\eqref{DefinitionOfLSFunctions}), $\mathbf{f}_k$ the vector of function coefficients for user $k$ (cf.~\eqref{DefinitionOfLSFunctions}), and $\mathbf{w}$ the vector of output files.
We also have
\begin{align}
\mathbf{e}_{n} &\triangleq [e_{n,1},e_{n,2},\hdots, e_{n,L}]^\intercal,\: n \in [N], \label{encoding-vectors-per-shot}\\
     \mathbf{z}& \triangleq [z_{1}, z_{2},\hdots, z_{N}]^ \intercal\label{linear-combinations}
\end{align}
respectively representing the encoding vector at server $n$, and the overall transmitted vector across all the servers (cf.~\eqref{EncodedFiles}).
Furthermore, we have
\begin{align}
     \mathbf{d}_{k} &\triangleq [d_{k,1},d_{k,2},\hdots, d_{k,N}]^ \intercal,\: k \in [K], \label{decoding-vectors-per-shot-1}\\
      \mathbf{f}'&\triangleq [F'_1,F'_2,\hdots,F'_K]^{\intercal}
\end{align}
respectively representing the decoding vector at user $k$, and the vector of the decoded functions across all the users.
In addition, we have
\begin{align}
      \mathbf{F}& \triangleq [\mathbf{f}_1,\mathbf{f}_2,\hdots,\mathbf{f}_K]^{\intercal} \in \F^{K \times L}, \label{Demand-Matrix-1}\\
        \mathbf{E} &\triangleq [\mathbf{e}_{1},\mathbf{e}_{2},\hdots, \mathbf{e}_{N}]^\intercal \in \F^{N \times L},\label{encoding-vectors-per-shot-1}\\
    \mathbf{D} &\triangleq [\mathbf{d}_1,\mathbf{d}_2, \hdots , \mathbf{d}_K]^{\intercal} \in \F^{K \times N} \label{DecodingMatrix-1}
\end{align}
where $\mathbf{F}$ represents the $K\times L$ matrix of all function coefficients across all the users, where $\mathbf{E}$ represents the $N\times L$ \emph{computation and encoding matrix} across all the servers, and where $\mathbf{D}$ represents the $K\times N$ \emph{communication and decoding matrix} across all the users.
We will henceforth assume that the columns of $\mathbf{F}$ are different from each other. 

Directly from~\eqref{DefinitionOfLSFunctions}, we have that
\begin{align}
    \mathbf{f} =[\mathbf{f}_1,\mathbf{f}_2,\hdots,\mathbf{f}_K]^{\intercal} \mathbf{w}\label{Functions-one}
\end{align}
and from \eqref{EncodedFiles} we have the overall transmitted vector to be 
\begin{align}
    \mathbf{z} =[\mathbf{e}_{1}, \mathbf{e}_{2}, \hdots,\mathbf{e}_{N}]^\intercal \mathbf{w} = \mathbf{E} \mathbf{w}. \label{EncodedCashedData-1}
\end{align}
Furthermore, directly from~\eqref{DecedFiles} we have that
\begin{align}
    F'_k= \mathbf{d}_{k}^{T} \mathbf{z}
\end{align}
and thus we have
\begin{align}
    \mathbf{f}' = [\mathbf{d}_1,\mathbf{d}_2,\hdots,\mathbf{d}_K]^{\intercal} \mathbf{z} = \mathbf{D}\mathbf{z}.\label{DecodedData-one}
\end{align}
Recall that we must guarantee that
\begin{align}
    \mathbf{f}'=\mathbf{f}\label{feasibility-one}.
\end{align}
After substituting \eqref{Functions-one}, \eqref{EncodedCashedData-1} and \eqref{DecodedData-one} into \eqref{feasibility-one}, we see that the feasibility condition in~\eqref{feasibility-one} is satisfied if and only if
\begin{align}
    \mathbf{D}\mathbf{E}\mathbf{w} = \mathbf{F}\mathbf{w}.\label{MainEquationWithW}
\end{align}
For this to hold for any $\mathbf{w}$, we must thus guarantee
\begin{align}
    \mathbf{D}\mathbf{E} = \mathbf{F}.\label{MainEquation}
\end{align}

At this point, we note that $\W_{n} = \supp(\mathbf{E}({n}, :))$ and $|\W_{n}| = \omega(\mathbf{E}({n}, :))$, which gives
\begin{align}
     \Lambda = \max_{n \in [N]} |\W_{n}| =  \underset{n \in [N]}{\max}\: \omega(\mathbf{E}({n}, :))  \label{delay}
\end{align}
revealing how the computational delay $\Lambda $ of our system is captured by the maximum number of non-zero elements of any row of $\mathbf{E}$. This is one of the two sparsity constraints that interest us. The other sparsity constraint can be seen by recalling that $|\W_{n}| = \omega(\mathbf{E}({n}, :))$ which gives
\begin{align}
    \Gamma = \sum^{N}_{n =1} |\W_n| = \omega(\mathbf{E}) \label{total-computation}
\end{align}
representing the total number of non-zero elements of $\mathbf{E}$, and which relates naturally to the cumulative computational cost across all servers.
Thus, being able to decompose $\mathbf{F}$ as $\mathbf{F}= \mathbf{D}\mathbf{E}$ where $\mathbf{E}$ corresponds to a reduced $\Lambda$ and $\Gamma$, allows us indeed to reduce the two computation costs. To do this, we will turn to perfect codes, as well as to quasi-perfect codes.

Note that whenever we say the multi-user linearly-decomposable is implemented based on the decomposition $\mathbf{D}\mathbf{E}=\mathbf{F}$, it means that based on the dimensionality $K, N$, we pick a code and choose its parity-check matrix as our $\mathbf{D}$, and then perform a syndrome decoding algorithm by regarding each column of $\mathbf{F}$ as a syndrome and each column of $\mathbf{E}$ as its corresponding error vector.

Before doing so, let us here provide a simple example to help clarify the setting and the notation.
\subsection{Example}
\label{Motivating-Example}
We consider the example of a system with a master node, $N=11$ servers, $K=5$ users, $L=12$ subfunctions, and a field of size $q=3$.
In our example, the jobs are defined (cf.~\eqref{Demand-Matrix-1}) by a demand  matrix that here takes the form
\begin{align*}
    \mathbf{F} =&
\left[ \begin{array}{cccccccccccccccccc}
 2   &   1  &    1  &    1   &   1    &  1   &   1    &  1   &   2   &   1   &   2    &  0   \\
     1   &   0   &   0    &  2    &  2   &   2   &   0     &  1    &  1   &   1   &   0    &  1   \\
     1    &  2   &   1  &    0   &   1    &  0   &   2    &  1  &    2  &    0   &   1   &   1   \\
     0     & 2    &  0   &   2    &  0    &  1     & 2     & 1    &  0    &  1    &  2   &   1   \\
     0      & 0    &  0   &   1    &  1    &  2     & 2     & 0    &  1    &  1    &  1   &   2  \\
\end{array}\right].
\end{align*}
In the computation and encoding phase, the master allocates the computation of the different subfunctions $f_1(.),f_2(.),\hdots,f_{12}(.)$ across the $11$ servers according to
\begin{align*}
    \mathcal{S}_1 &=\{ 1,8,10\},\: \mathcal{S}_2 = \{2,6\},\:  \mathcal{S}_3 = \{5,9\}, \\
    \mathcal{S}_4 &= \{4\}, \: \mathcal{S}_5 = \{7,11\}, \: \mathcal{S}_6 =\{1,6,12\}, \: \mathcal{S}_7 =\{3\}, \\ 
    \mathcal{S}_8 &=\{2\}, \: \mathcal{S}_9 =\{3,10\}, \: \mathcal{S}_{10} =\{7,12\}, \: \mathcal{S}_{11} =\{5,8\}
\end{align*}
forcing  the two costs to be $\Gamma = \sum^{N}_{n =1} |\W_n| = 21$ and $\Lambda = \max_{n \in [N]} |\W_{n}| =3$. 
After computing their designated output files, each server $n$ transmits $z_n$
corresponding to a computation and encoding matrix (cf.~\eqref{EncodedCashedData-1}) which here takes the value
\vspace{10pt}
\begin{align*}
    \mathbf{E} =&
    \left[ \begin{array}{cccccccccccccccccc}
     2  &   0  &  0  &   0   &  0   &  0  &   0    & 1  &   0   &  1  &   0  &   0 \\
     0  &   1   &  0   &  0   &  0   &  1    & 0   &  0  &   0  &   0  &   0   &  0 \\
     0  &   0  &   0  &   0  &   1  &   0  &   0   &  0  &   2   &  0   &  0   &  0 \\
     0   &  0   &  0   &  2   &  0  &   0   &  0   &  0  &   0    & 0   &  0   &  0 \\
     0  &   0   &  0  &   0  &   0  &   0   &  2  &   0  &   0  &   0   &  1   &  0 \\
     1   &  0   &  0   &  0  &   0  &   2   &  0  &   0  &   0   &  0   &  0   &  2  \\
     0   &  0   &  1   &  0  &   0   &  0  &   0  &   0   &  0   &  0   &  0   &  0   \\
     0   &  2  &   0    & 0   &  0  &   0  &   0   &  0   &  0   &  0   &  0   &  0   \\
     0   &  0   &  1   &  0   &  0  &   0   &  0   &  0  &   0   &  2   &  0   &  0  \\
     0   &  0   &  0   &  0  &   0   &  0   &  1   &  0  &   0   &  0    & 0   &  2 \\
     0   &  0   &  0   &  0   &  2  &   0   &  0   &  2   &  0   &  0    & 0   &  0  \\
\end{array}\right]
\end{align*}
and which indeed abides by the constraint $\underset{n \in [N]}{\max}\: 
 \omega(\mathbf{E}({n}, :)) = \Lambda = 3$ (cf.~\eqref{delay}) and the constraint $\omega(\mathbf{E}) = \Gamma =  21$ (cf.~\eqref{total-computation}).

Subsequently, the master asks each server $n$ to send its generated $z_{n}$ to its designated receiving users, where for each server these user-sets are
\begin{align*}
    \T_{1} &= \{1,2,\hdots, 5\}, \: \T_{2} = \{1,2,\hdots, 4\},\:\\ \T_{3} &= \{1,2,3,5\}, \: \T_{4} =\{1,2,4,5\},\:\\ \T_{5} &=\{1,3,4,5\},\: \T_{6} = \{2,3,4,5\}, \: \T_{7} = \{1\}, \T_{8} = \{2\},\:\\
    \T_{9} &= \{3\},\: \T_{10} = \{4\}, \T_{11} = \{ 5\}
\end{align*}
which simply says that, for example, server 2 will broadcast $z_{2}$ to users $1,2,3,4$. Subsequently, the decoding procedure is executed, 
adhering to a decoding matrix which takes the value
\begin{align}
    \mathbf{D}  =
    \left[ \begin{array}{ccccccccccc}
  1  &   1    & 1  &   2    & 2   &  0    & 1    & 0   &  0   &  0   &  0\\
     1 &    1   &  2 &    1  &   0    & 2  &   0   &  1   &  0  &   0&     0\\
     1  &   2   &  1  &   0   &  1   &  2  &   0  &   0   &  1   &  0 &    0\\
     1    & 2  &   0   &  1  &   2   &  1   &  0  &   0  &   0 &    1  &   0\\
     1    & 0  &   2    & 2  &   1   &  1  &   0 &    0   &  0   &  0   &  1
\end{array}\right].\label{decoding-golay}
\end{align}
In the next section, we will show that the decoding matrix $\mathbf{D}$, is drawn from a class of perfect codes whose properties --- as we show below --- allow for an improved performance. 
\section{Main Results} \label{results}
We proceed with the main results of our work.
\begin{Theorem}\label{m-1}
The optimal computational delay $\Lambda$ of the $(K,N)$ multi-user linearly decomposable problem implemented based on the decomposition $\mathbf{DE} = \mathbf{F}$, is bounded as 
\begin{align}
        \Lambda \leq \min\{L,\sum^{\tau}_{i=1} { N-1 \choose i-1} (q-1)^{i} + (1-\mu_{\tau}) q^{K}\} \label{r1}
        \end{align}
where $\tau$ and $\mu_\tau $ are respectively the packing radius and the corresponding packing density of $\mathcal{C}_{\mathbf{D}}$.
\end{Theorem} 
\begin{proof}
    Let us rewrite \eqref{MainEquation} as
    \begin{align}
        \mathbf{D} \mathbf{E}(:,l) = \mathbf{F}(:,l), \: \forall l \in [L]\label{maineq}
    \end{align}
    and let us note that given a certain $\mathbf{D}$, we ask that for every column $\mathbf{F}(:,l) \in \F^{K}$, the resulting $\mathbf{E}(:,l)$ has a reduced number of non-zero elements. To achieve this, after regarding $\mathbf{D}$ to be a parity-check matrix of a linear code, we employ a syndrome decoder, rendering the `error pattern' $\mathbf{E}(:,l)$ to be a coset leader associated to syndrome $\mathbf{F}(:,l)$. Now let's define  the following set
    \begin{align}
        \mathcal{S} \triangleq  \{l: \omega(\mathbf{E}(:,l)) \leq \tau\} \label{Good-set-Definition}
    \end{align}
    which represents the syndromes $\mathbf{F}(:,l)$  for which the weight of their associated coset leader is below the packing radius $\tau$. From the basic error-correcting argument, we know that all the error patterns having no more than than $\tau$ non-zero elements are indeed present in $\mathcal{S}$, and thus we have that 
    $|\mathcal{S}|\leq  \sum^{\tau}_{i=1} { N \choose i}(q-1)^{i}$.
    Let us now focus on a row subvector $\mathbf{E}(n,\mathcal{S})$, and let us count the number of its non-zero elements. Since we know that for all $l \in \mathcal{S}$ it is indeed the case that $\omega(\mathbf{E}(:,l))  \leq \tau$, and since the collection of all such columns $\mathbf{E}(:,l)$ contains all possible error patterns of weight up to $\tau$, we can conclude that 
    \begin{align}
        \omega(\mathbf{E}(n,\mathcal{S})) \leq  \sum^{\tau}_{i=1} { N-1 \choose i-1}(q-1)^{i},\:\:\: \forall n \in [N] \label{l-bound}
    \end{align}
    because i) each element $\mathbf{E}(n,l) \neq 0, l \in \mathcal{S} $ can take any one of $(q-1)$ possible values, because ii) such a column vector $\mathbf{E}(:,l)$ (where again $l \in \mathcal{S}$) has at most $\tau$ such non-zero elements, and because iii) there exist at most ${ N-1 \choose i-1} (q-1)^{i-1}$ vectors $\mathbf{E}(:,l) \neq 0, l \in \mathcal{S} $ whose $n$th entry $\mathbf{E}(n,l)$ is non-zero and whose total weight is $i\leq \tau$. 
    Summing across all $i = 1,2,...,\tau$ yields the first term in our bound in~\eqref{l-bound}. 
    To conclude the proof, we proceed to count the number of syndromes (columns $\mathbf{F}(:,l)$ of $\mathbf{F}$) whose coset leaders (referring to the corresponding column $\mathbf{E}(:,l)$ of $\mathbf{E}$) have weight strictly bigger than $\tau$. To do so, we first note that the number of points that are not covered by the ball $\mathcal{B}(\mathbf{c},\tau),\: \mathbf{c} \in \mathcal{C}_{\mathbf{D}}$, is $(1 - \mu_{\tau})q^{N}$. Recalling that each coset corresponds to $q^{N-K}$ points (vectors) in $\mathbf{GF}(q)^{N}$, we can conclude that the number of cosets in question is $(1 - \mu_{\tau})q^{K}$. These are the cosets corresponding to syndromes with indices from $\mathcal{S}'\triangleq \{l: \omega(\mathbf{E}(:,l)) > \tau\}$.     Thus, for the worst-case scenario of interest, we can consider that for all $l$ for which $\omega(\mathbf{E}(:,l) )>\tau$, it is the case that we will encounter a non-zero $\mathbf{E}(n,l)$ for any  $n\in [N]$, which is reflected in the addition of the second term $(1-\mu_{\tau}) q^{K}$ in our bound. Naturally, $L$ is a trivial upper bound on $\Lambda$.  This concludes the proof of the theorem. 
\end{proof}
We now proceed with the second result, this time regarding the cumulative computation cost. 

\begin{Theorem}\label{m-2}
The optimal cumulative computation cost $\Gamma$ of the $(K,N)$ multi-user linearly decomposable problem implemented based on the decomposition $\mathbf{DE} = \mathbf{F}$, is bounded as
    \begin{align}
        \Gamma  \leq   \min\{NL,\sum^{\tau}_{i=1} { N \choose i}(q-1)^i i+ (1-\mu_{\tau}) q^K   \rho \}\label{r2}
        \end{align}
where $\tau, \rho$ and $\mu_\tau $ are respectively the packing radius, covering radius, and packing density of $\mathcal{C}_{\mathbf{D}}$.
\end{Theorem}
\begin{proof} 
The proof follows the steps of the proof of Theorem \ref{m-1}, up until the definition in~\eqref{Good-set-Definition}, where similarly we now know that all the error patterns having no more than $\tau$ non-zero elements are present in the set of all possible $\mathbf{F}(:,l),\:\: l \in \mathcal{S}$, and thus we know that the sum of the Hamming weights of the coset leaders corresponding to the syndromes $\mathbf{F}(:,l),\:\: l \in \mathcal{S}$, takes the form $\sum^{\tau}_{i=1} { N \choose i} i (q-1)^i $  which matches the first term of our bound. Regarding the second term, we know that for any $l \notin \mathcal{S}$, the number of non-zero elements in $\mathbf{E}(:,l)$ is --- by definition of the covering radius --- no greater than $\rho$. We also see, following the ball arguments in the proof of the previous theorem, that the set $\mathcal{S}'\triangleq \{l: \omega(\mathbf{E}(:,l)) > \tau\}$ has size $|\mathcal{S}'|= (1 - \mu_{\tau})q^K$. Hence, we can now conclude that the sum of the weights of the coset leaders of the vectors (seen as syndromes) in $\mathcal{S}'$, is upper bounded by $\rho |\mathcal{S'}|$. 
\end{proof}

\subsection{The Connection to Perfect and Quasi-Perfect Codes}
At this point, we note that the above bound on $\Gamma$ is reduced when the covering radius $\rho$ is reduced, while the bound on $\Lambda$ is reduced when, for any given $\tau$, the packing density $\mu_{\tau}$ is increased. For this reason we are looking for codes (whose parity check matrix will be used as the decoding matrix $\mathbf{D}$) that indeed minimize $\rho$ and increase $\mu_{\tau}$ for a fixed $\tau$. This special and rare property sought in $\mathcal{C}_{\mathbf{D}}$ is indeed attributed to the well-known class of perfect codes~\cite{etzion2022perfect}. 

It is though known (cf.~\cite{tietavainen1973nonexistence}) that there exist few such perfect codes, for a few select dimensions. Thus, for other dimensions, we will resort to the use of quasi-perfect codes which indeed ensure a similar performance, by guaranteeing optimal $\rho$ and near-optimal $\mu_\tau$. We elaborate on this later on. 


\begin{remark}
Looking back to our previous example corresponding to $N=11, K=5, L=12$ and $q=3$, the distributed computing solutions employed an $\mathbf{E}$ matrix that resulted from an optimal syndrome decoding process of a code whose parity check matrix $\mathbf{D}$ is indeed that of a ternary Golay code, which is a known perfect code.
\end{remark}

\subsection{The Special Case of Maximal Basis Set}
We here consider also the case where $L = q^K$. This particular case of having a maximum number of basis subfunctions presents interesting advantages. In essence, under the same assumptions as before, we are now able to set the decoding matrix $\mathbf{D}$ once, well before the desired functions are declared. This allows us to have a fixed network connectivity, where --- under the assumption of knowing in advance $K,N$ --- we can fix $\mathbf{D}$, and we can then  account for each $\mathbf{F}$ simply by altering $\mathbf{E}$. For this setting, we have the following propositions. 

\begin{Proposition}\label{m-4}
The optimal computational delay $\Lambda$ and cumulative computation cost $\Gamma$ of the $(K,N)$ multi-user linearly decomposable problem with maximal basis, are lower bounded as 
\begin{align} \label{LowerBoundsNew}
        \Lambda &\geq \sum^{\tau}_{i=1} { N-1 \choose i-1} (q-1)^{i}, \ \ \Gamma \geq \sum^{\tau}_{i=1} { N \choose i}(q-1)^{i} i
        \end{align}
where $\tau$ is the packing radius of $\mathcal{C}_{\mathbf{D}}$. 
\end{Proposition} 
\begin{proof}
    Regarding the bound on $\Lambda$, we follow all the steps of the proof in Theorem~\ref{m-1}, with the only differences being firstly that the used inequality $|\mathcal{S}| \leq \sum^{\tau}_{i=1} { N \choose i}(q-1)^{i}$ is now automatically forced into an equality, and secondly that we consider, by choice, a lower bound $|\mathcal{S}'|\geq 0$. 
    Exactly the same approach applies for the bound on $\Gamma$.    
\end{proof}
Regarding optimality, we have the following proposition.
\begin{Proposition}\label{m-5}
The optimal computational costs $\Lambda$ and $\Gamma$ for the cases $(K,N)$ for which a perfect code exists, take the form 
\begin{align}
        \Lambda = \sum^{\tau}_{i=1} { N-1 \choose i-1} (q-1)^{i},  \ \ \Gamma = \sum^{\tau}_{i=1} { N \choose i}(q-1)^{i} i
        \end{align}
where $\tau$ is the packing radius of the used perfect code $\mathcal{C}_{\mathbf{D}}$. 
\end{Proposition} 
\begin{proof}
For the case where $(K,N)$ accepts a perfect code, we know (cf.~\cite{etzion2022perfect}) that $\mu_\tau = 1$ for which the upper bounds in \eqref{r1} and \eqref{r2} match the corresponding upper bounds in~\eqref{LowerBoundsNew}. 
\end{proof}
We also have the following proposition for a much broader range of dimensionalities. 
\begin{Proposition} 
For all cases $(K,N)$ for which there exists a quasi-perfect code $\mathcal{C}_{\mathbf{D}}$, the optimal performance is upper bounded as $\Lambda < \sum^{\tau}_{i=1} { N-1 \choose i-1} (q-1)^{i} + { N \choose \tau +1}(q-1)^{\tau+1}$ and $\Gamma <  \sum^{\tau+1}_{i=1} { N \choose i}(q-1)^i i$, where $\tau$ is the packing radius of the corresponding quasi-perfect code $\mathcal{C}_{\mathbf{D}}$. 
\end{Proposition} 
\begin{proof}
The proof is direct by noting that quasi-perfect codes have $\mu_\tau > 1- { N \choose \tau +1}(q-1)^{\tau+1}q^{-K}$ and $\tau = \rho -1$ \cite{etzion2022perfect}. 
\end{proof}
\begin{remark}
Given the non-existence results in \cite{tietavainen1973nonexistence}, it is in fact not difficult to show that quasi-perfect codes minimize the gaps between the upper bounds in \eqref{r1} and \eqref{r2} and the corresponding lower bounds in \eqref{LowerBoundsNew}.
\end{remark}

\section{conclusion}\label{conclusion}
We have explored the computational cost of the multi-user distributed computing setting of linearly decomposable functions, which nicely captures various problems such as the distributed gradient coding problem \cite{tandon2017gradient}, the distributed linear transform problem \cite{dutta2016short}, the distributed matrix multiplication problem, and the distributed multivariate polynomial computation problems \cite{yu2020straggler,jia2021cross}, among others.  The work established various upper and lower bounds on the computational delay $\Lambda$ and the cumulative computation cost $\Gamma\in[0,NL]$, and revealed new connections to the packing and covering capabilities of codes thus revealing for the first time powerful connections with the class of perfect codes. 

\bibliographystyle{ieeetr}
\bibliography{ref}
\end{document}